# Nouvelle analyse des phénomènes vibratoires en tournage

# New vibrations phenomena analysis in turning

Claudiu-Florinel BISU[a, 1-2-3], Philippe DARNIS[b,2], Jean-Yves K'NEVEZ[b, 1], Olivier CAHUC[c, 1], Raynald LAHEURTE[b, 1-2], Alain GÉRARD[d, 1], Constantin ISPAS[d,3]

[1] : Université Bordeaux 1 – CNRS UMR 5469
Laboratoire de Mécanique Physique
351, Cours de la libération   33405 Talence Cedex – France (UE)
Fax.: +(33) (0)5 40 00 69 64

E-mail: jean-yves.knevez@u-bordeaux1.fr   / Tél.:+ (33) (0)5 40 00 89 28
E-mail: olivier.cahuc@u-bordeaux1.fr   / Tél.:+ (33) (0)5 40 00 87 89
E-mail: raynald.laheurte@u-bordeaux1.fr   / Tél.:+ (33) (0)5 40 00 38 47
E-mail: alain.gerard@u-bordeaux1.fr   / Tél.:+ (33) (0)5 40 00 62 23

[2] : Université Bordeaux 1 – IUT EA 496
Laboratoire de Génie Mécanique et Matériaux de Bordeaux
Domaine Universitaire
15, rue Naudet   33175 Gradignan Cedex – France (UE)
Fax. :+(33) (0)5 56 84 58 43

E-mail: philippe.darnis@u-bordeaux1.fr  / Tél.:+ (33) (0)5 56 84 79 76

[3] : University Politehnica from Bucharest,
Machines and Production Systems Department
Splaiul Independentei, 313   Bucharest 060042 – Roumanie (UE).
Fax : +(40) 214 104 267

E-mail: cfbisu@gmail.com /  Tél.: +(33) 5 40 00 38 58

a : Doctorant
b : Maître de Conférences
c : Maître de Conférences Habilité à Diriger les Recherches
d : Professeur des Universités

**Correspondance à adresser à** : **Alain GERARD** Professeur Université Bordeaux 1
Mél : *alain.gerard@u-bordeaux1.fr*
Tél : 05 40 00 62 23

**Résumé**

Dans le processus de fabrication par enlèvement de matière sur machines-outils, l'apparition des vibrations est inévitable. Dans les situations où l'amplitude dépasse les limites de la précision dimensionnelle, le phénomène vibratoire est préjudiciable. L'étude expérimentale présentée ici permet d'identifier les principaux paramètres relatifs au comportement dynamique du système usinant. La localisation des déplacements de la pointe de l'outil dans un plan est démontrée grâce aux résultats expérimentaux. L'existence de ce plan et les corrélations avec les caractéristiques élastiques du système usinant permettent de simplifier le modèle dynamique 3D. Avec l'approche expérimentale réalisée, un modèle dynamique est proposé en concordance avec les résultats expérimentaux.

**Mots clés :** vibrations / modèle expérimental / plan des déplacements / vibrations auto-entretenues

**Abstract**

In the cutting process, machine-tools vibrations are generally a real problem when the amplitude crosses the limits of dimensional or surface quality workpiece precision required. It is necessary to develop models taking into account the three-dimensional vibratory approach to control the vibrations phenomena. An experimental study has been realized to understand the behavior of the cutting system and to identify its vibrations properties. The displacement's localization of the tool in a spatial plan, showed by the experimental results, allows us to simplify the three-dimensional dynamical model. In this study, the experimental approach is completely presented and the first points of the three dimensional vibratory model are proposed according to the experimental results.

# 1 Introduction

L'usinage par enlèvement de matière est la principale technique utilisée dans la production des pièces mécaniques pour l'industrie automobile, aéronautique, ferroviaire etc. Il représente une partie importante du coût des pièces. L'introduction de nouvelles générations de machines travaillant à grande vitesse de coupe et à grande vitesse d'avance, le développement de nouveaux outils et la nécessité d'une plus grande productivité associée à une meilleure qualité des produits élaborés nécessitent une connaissance approfondie des processus d'usinage [1]. De mauvais choix des conditions de coupe peuvent provoquer, lors de l'usinage, des vibrations de l'outil, se transmettant au porte-outil et à la machine. Ces vibrations constituent un obstacle majeur pour la recherche d'une plus grande productivité et d'une meilleure qualité des pièces réalisées. Les vibrations accélèrent l'usure, la dégradation de l'outil et conduisent à des états de surface médiocres. De plus, elles risquent aussi d'engendrer des dommages, voire même des ruptures, dans les composants des machines.

Néanmoins, dans le processus d'enlèvement de matière sur machines-outils l'apparition de vibrations est inévitable. Dès que l'amplitude dépasse les limites de la précision dimensionnelle et selon la forme des surfaces génératrices, le phénomène vibratoire est préjudiciable. Les vibrations à l'interface outil/copeau, lors de l'usinage sont principalement dues aux variations du frottement à l'interface, au contact sur la face en dépouille de l'outil et aux variations d'épaisseur et de largeur usinées. Ces dernières proviennent de la génération d'une surface ondulée lors de la passe précédente, qui influence le comportement dynamique de l'ensemble outil/porte-outil lors de la passe suivante [1]. Le mouvement de l'outil est alors entretenu. La fluctuation des efforts de coupe excite le système. Ces mécanismes se produisent simultanément, sont interdépendants et sont à l'origine des vibrations auto-entretenues.

Ainsi, il est nécessaire de développer des modèles permettant d'étudier les phénomènes vibratoires rencontrés au cours de l'usinage et de prévoir les conditions de stabilité des processus de coupe. Le modèle dynamique 3D réalisé est basé sur un modèle semi-analytique tridimensionnel de la coupe développé par deux laboratoires bordelais [2].

Dans la littérature, les modèles de type Merchant servent généralement de base pour modéliser les efforts dynamiques lors de la coupe. Par contre, ces modèles prennent rarement en compte la configuration globale pièce/outil/machine (**POM**). L'originalité des travaux présentés ici concerne l'intégration, dans le modèle, de l'influence de la géométrie de l'outil et de son déplacement pendant l'usinage ainsi que des évolutions des conditions de contact pièce/outil/copeau.

Une étude expérimentale préliminaire permet d'identifier les principaux paramètres relatifs au comportement dynamique du système usinant. La localisation des déplacements de la pointe de l'outil dans un plan est démontrée grâce aux résultats expérimentaux. L'existence de ce plan et les corrélations avec les caractéristiques élastiques du système usinant permettent de valider le dispositif expérimental et de simplifier le modèle dynamique 3D (Fig. 1).

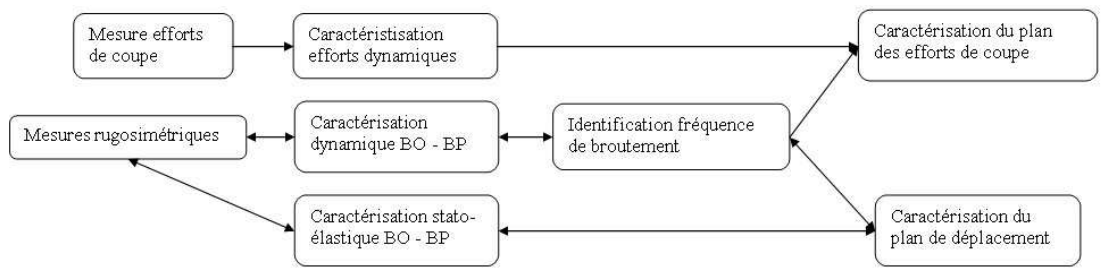

**Fig. 1**. Synoptique de la démarche adoptée.

Cet article comprend trois parties. La première est consacrée au dispositif expérimental et à la caractérisation du système usinant. La seconde présente l'analyse des résultats expérimentaux, met en évidence l'existence d'un plan des déplacements de la pointe de l'outil lors de l'usinage, et permet la validation du dispositif expérimental. La troisième partie est consacrée aux conséquences de ces résultats sur l'écriture d'un modèle semi-analytique prenant en compte les vibrations.

## 2 Dispositif expérimental

Les essais de coupe sont réalisés sur un tour conventionnel (Fig. 2a). Le comportement dynamique est identifié à l'aide d'un accéléromètre 3D fixé sur l'outil et de deux accéléromètres 1D positionnés sur le tour, côté broche. Les efforts et les moments à la pointe de l'outil sont obtenus à l'aide d'un dynamomètre à six composantes [3]. La vitesse instantanée de la pièce est donnée par un codeur rotatif.

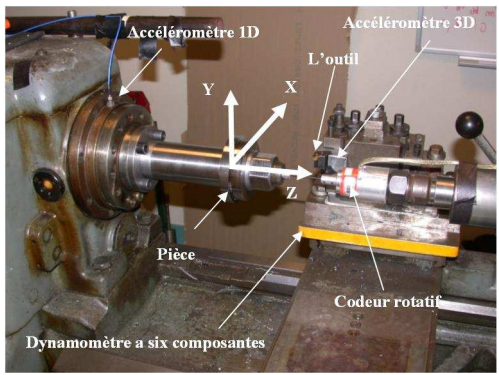
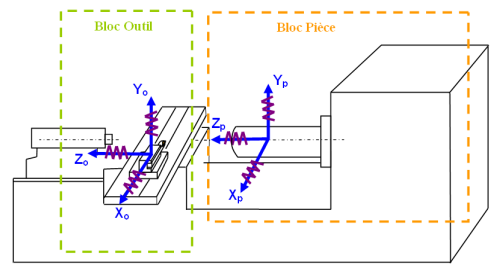

**Fig. 2a**. Dispositif expérimental de mesure des vibrations.

**Fig. 2b**. Décomposition du système.

## 3 Caractérisation du système usinant

Afin de caractériser le système usinant, nous réalisons une analyse du comportement élastique du système pour déterminer le mouvement de l'outil. Les déplacements de l'outil sont alors proportionnels à la raideur du système.

Le système usinant est divisé en deux blocs (Fig. 2b): un bloc outil (**BO**) et un bloc pièce (**BP**). Sa caractérisation est effectuée en trois étapes : une analyse statique, suivie d'une analyse modale par impact, et terminée par une analyse dynamique.

## 3.1 Analyse statique

L'étude statique a pour but de caractériser les raideurs équivalentes statiques pour déterminer le comportement élastique tridimensionnel du système usinant. Les essais statiques consistent à charger (chargement effectué à l'aide d'un système vis écrou et contrôlé par un capteur de force suivant x, y, z) et à mesurer le torseur des petits déplacements (6 capteurs de micro déplacement unidirectionnels suivant figure 3) des différents blocs **BO** et **BP** suivant trois directions.

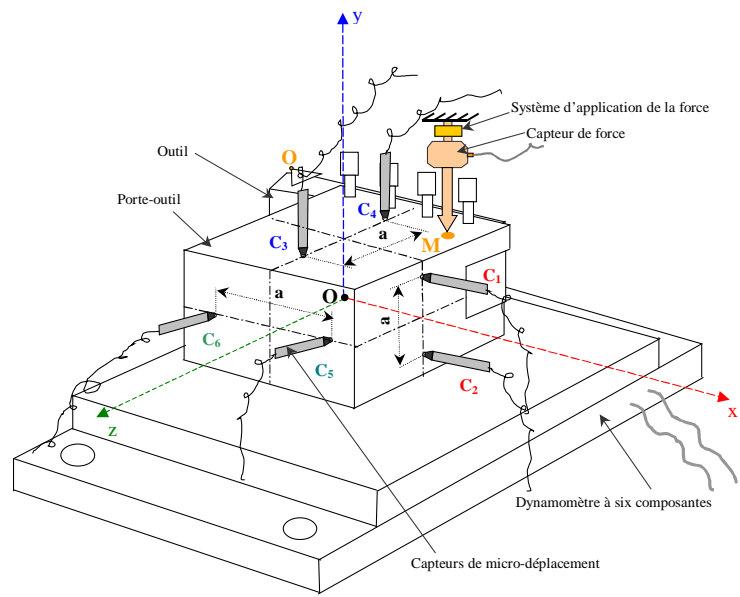

**Fig. 3.** Protocole expérimental pour la caractérisation statique du bloc outil (**BO**).

Ce dispositif expérimental permet de déterminer la matrice de raideur globale [$K_{BO}$] et [$K_{BP}$] respectivement des systèmes **BO** et **BP** exprimées au point de contact A de la pièce et de l'outil. La forme générale est donnée par l'expression :

$$[T] = [K] \cdot [D] \quad (1)$$

où $[T]$ représente la matrice des torseurs des actions mécaniques appliqués, $[D]$ la matrice des torseurs des petits déplacements associés [3], tandis que la matrice [$K$] est donnée par.

$$\begin{bmatrix} K_i \end{bmatrix}_{6\times 6 \; A,XYZ_i} = \begin{bmatrix} K_{iF} & K_{iFC} \\ {}_{3\times 3} & {}_{3\times 3} \\ K_{iCF} & K_{iC} \\ {}_{3\times 3} & {}_{3\times 3} \end{bmatrix}_{A,XYZ_i} \quad (2)$$

où $K_{iF}$, $K_{iC}$, $K_{iCF}$ et $K_{iFC}$ sont respectivement les matrices de raideurs de déplacements, de rotations et de couplage rotations / déplacements et déplacements / rotations exprimées en A avec $i = BO$ ou $i = BP$ pour les deux blocs figure 4 et figure 5.

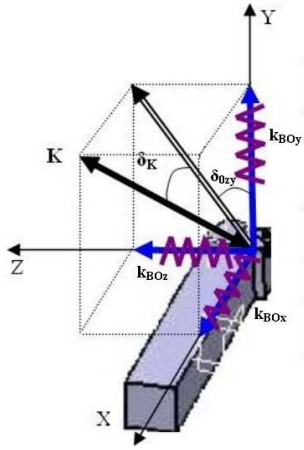 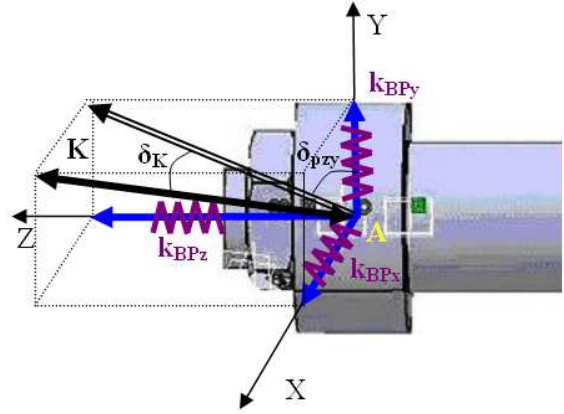

Fig.4. Raideurs du système **BO**.　　　　Fig.5. Raideurs du système **BP**.

Dans cette étude préliminaire, seule la matrice $K_{iF}$ sera considérée :

$$\begin{bmatrix} K_{iF} \end{bmatrix}_{3\times 3 \; A,XYZ_i} = \begin{bmatrix} k_{ix} & k_{iyx} & k_{izx} \\ k_{ixy} & k_{iy} & k_{izy} \\ k_{ixz} & k_{iyz} & k_{iz} \end{bmatrix}_{A,XYZ_i} . \quad (3)$$

Pour la partie **BP**, la matrice $K_{iF}$ est diagonale (conséquence des symétries, absence de frottement, … ). En revanche, il n'en est pas de même pour la partie **BO** du fait des assemblages et des frottements qui les accompagnent. Cependant, après assemblage des deux matrices et changement de base il vient :

$$\begin{bmatrix} K_F \end{bmatrix}_{3\times 3 \; A,XYZ} = \begin{bmatrix} 2.6\cdot 10^7 & 3.5\cdot 10^{-9} & 3.5\cdot 10^{-10} \\ 2.6\cdot 10^{-9} & 2.2\cdot 10^7 & 5.6\cdot 10^{-9} \\ -9\cdot 10^{-9} & -1\cdot 10^{-9} & 2.9\cdot 10^8 \end{bmatrix} . \quad (4)$$

Cette expression montre bien que les couplages sont négligeables. Aussi seuls les termes diagonaux seront pris en compte par la suite.

### 3.2 Analyse modale par impact

Le but des essais est d'identifier le comportement vibratoire du processus de coupe et le comportement dynamique de l'ensemble **POM**. Sur la base des informations expérimentales obtenues, nous déterminons et nous caractérisons la zone des vibrations auto-entretenues. Cette identification est possible par la superposition modale appliquée à l'outil, à la pièce et à la machine.

A l'aide d'un marteau à impact, les fréquences propres de chaque bloc sont déterminées selon les directions X, Y, Z. La figure 6a donne pour le cas de l'outil, la plage des fréquences déterminée par cette méthode. Les résultats obtenus sont présentés dans la figure 6b. La plage des fréquences propres est représentée en réalisant une superposition modale pour chaque élément composant le système. Ces résultats sont cohérents avec ceux rencontrés dans la littérature [4] à [5].

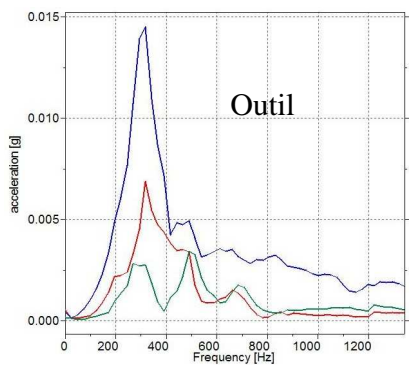
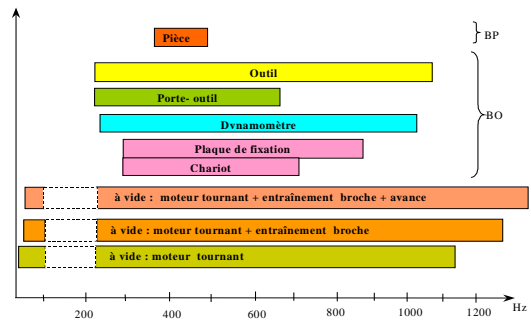

**Fig. 6a.** Exemple de spectre obtenu pour l'outil.   **Fig. 6b**. Superposition des plages de fréquences propres du système usinant pour chaque élément.

**3.3 Analyse dynamique**

La caractérisation dynamique du système usinant est complétée par une analyse suivant trois configurations. Pour appliquer la superposition modale, les trois configurations suivantes sont analysées: moteur électrique tournant, moteur électrique tournant avec entraînement de la broche, moteur électrique tournant avec broche et mouvement d'avance embrayé. Les fréquences prédominantes mesurées à l'aide d'accéléromètres 3D (sur le **BO**) et 1D (sur le **BP**) se situent aux environ de 525 Hz sur X, 855 Hz sur Y et 525 Hz sur Z pour chacune des trois configurations citées (Fig. 6b).

L'analyse modale du système sert ainsi de base à l'identification des vibrations auto-entretenues.

**4 Caractérisation du processus de coupe**

La méthode de caractérisation des vibrations auto-entretenues est issue des travaux [6]. Afin de déterminer la zone des vibrations auto-entretenues, les paramètres de coupe restent constants excepté la profondeur de passe (ap). L'apparition des vibrations auto-entretenues est clairement observée pour ap = 5mm. Autour de ce point de fonctionnement, des séries de tests sont effectuées, pour déterminer la plage des vibrations auto-entretenues, en maintenant la même valeur de ap et la même vitesse de rotation (N = 690 tr/min) de la pièce mais en faisant varier les avances (f) : f = 0.05 mm/tr, f = 0.0625 mm/tr, f = 0.075 mm/tr et f = 0.1 mm/tr. Pour ces essais, l'outil utilisé est de type TNMA 160412 (nuance carbure-SUMITOMO ELECTRIC) et le matériau usiné est de type 42CrMo4. Les éprouvettes (Fig. 5) sont d'un diamètre 120 mm et d'une longueur de 30 mm. La mesure des vibrations, effectuée pendant l'usinage grâce à un accéléromètre 3D, permet d'obtenir les résultats présentés à la figure 7. Les fréquences de vibrations de coupe se situent aux environs de 195 Hz pour les trois axes avec une amplitude prépondérante sur l'axe Y. Ces fréquences sont inférieures à celles relevées lors de la caractérisation du système usinant (Fig. 6b) ; elles correspondent donc bien aux vibrations auto-entretenues liées spécifiquement au phénomène de coupe.

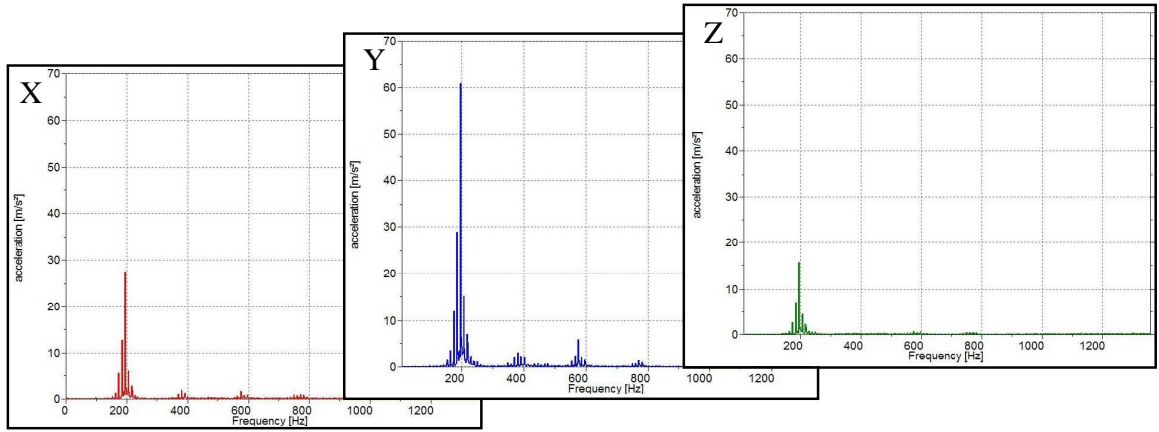

**Fig. 7.** Vibrations auto-entretenues suivant X, Y, Z ; ap = 5 mm, f = 0.1 mm/tr,

Vc = 238 m/min.

Le torseur des efforts est mesuré avec un dynamomètre à 6 composantes [3]. Par le calcul, nous déduisons les composantes des forces et des moments à la pointe de l'outil. Une analyse FFT du signal conduit à situer la fréquence d'excitation aux environs de 190 Hz. Ces résultats concordent avec ceux de la littérature [6] à [8] qui montrent que les forces varient aussi avec les fréquences des vibrations auto-entretenues.

Les vibrations auto-entretenues ont une influence sur la qualité de surface des pièces (Fig. 8). L'analyse FFT des données rugosimètriques montrent un pic de fréquence localisé autour de 200 Hz, cohérent avec les données accélérométriques.

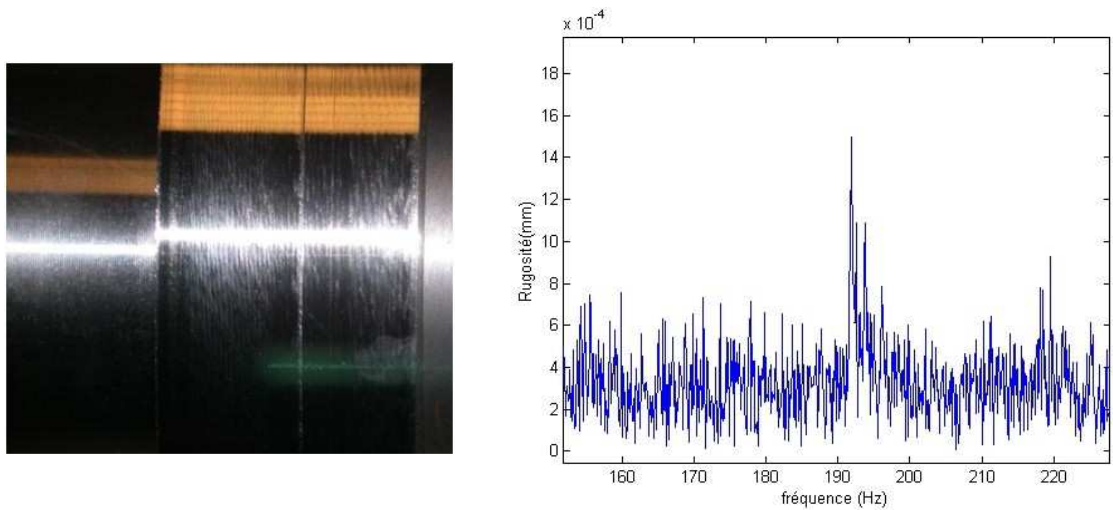

**Fig. 8.** Photo et FFT du profil de rugosité de la pièce usinée.

En conclusion le phénomène des vibrations auto-entretenues est bien identifié. La méthode de superposition des fréquences a permis la caractérisation du système usinant. L'identification des vibrations auto-entretenues est également corrélée à l'analyse de la surface usinée.

**5 Analyse du comportement du bloc outil**

L'analyse des données accélérométriques permet d'établir l'existence d'un plan des déplacements dans lequel la pointe de l'outil décrit une ellipse. De même, l'analyse des efforts met en évidence une évolution dans un plan de l'effort de coupe variable $Fv$ autour d'une valeur nominale $Fn$ (Fig. 9). Cet effort variable est un effort tournant qui génère les déplacements (u) de l'outil et entretient les vibrations du système élastique **BO** [6] à [7].

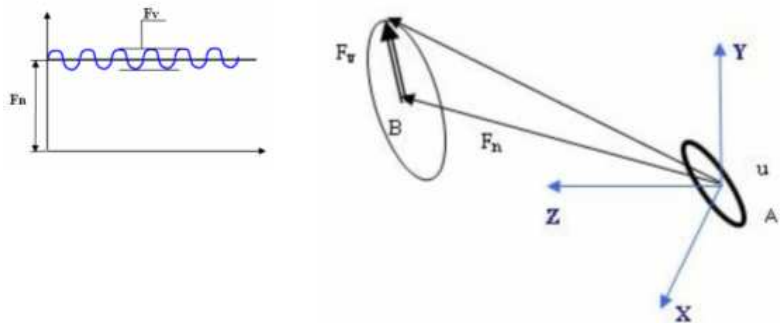

**Fig. 9.** Evolution dans un plan de l'effort de coupe variable *Fv* autour de sa valeur nominale *Fn*.

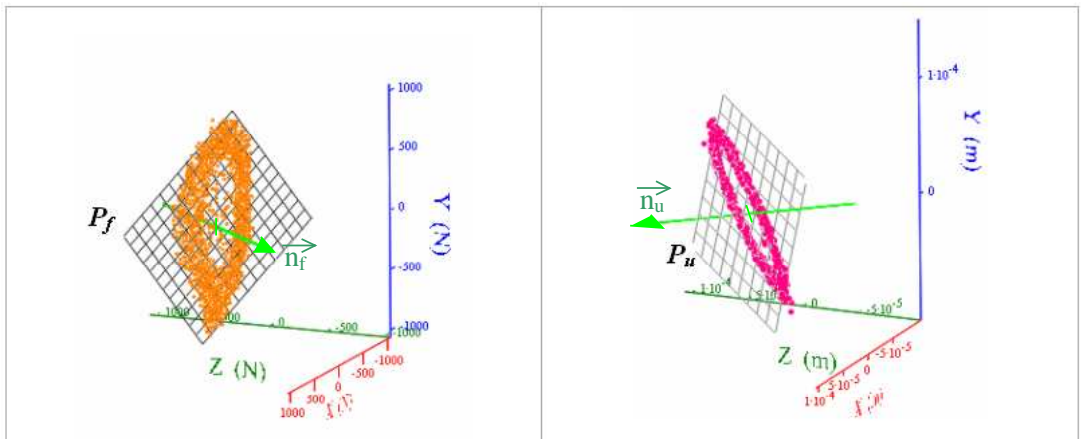

**Fig. 10.** Plan des forces variables et des déplacements dans le repère X, Y, Z.

Cas : ap= 5 mm, f = 0.1 mm/tr, Vc= 238 m/min.

Les projections du vecteur effort et du vecteur déplacement permettent de caractériser deux plans : $P_f$ et $P_u$. La position de ces plans dans le repère global X, Y, Z est déterminée à l'aide de leurs normales respectives $\vec{n_f}$ et $\vec{n_u}$ dans ce repère. Notons que la meilleure corrélation a lieu suivant l'axe de coupe Y tableau 1.

**Tableau 1.** Normales ($\vec{n}_f, \vec{n}_u$) aux plans des déplacements et des forces suivant les trois directions de coupe, en fonction de l'avance.

| f (mm/tr) | 0.05 | | 0.0625 | | 0.075 | | 0.1 | |
|---|---|---|---|---|---|---|---|---|
| Normale | $\vec{n}_f$ | $\vec{n}_u$ | $\vec{n}_f$ | $\vec{n}_u$ | $\vec{n}_f$ | $\vec{n}_u$ | $\vec{n}_f$ | $\vec{n}_u$ |
| Suivant X | 0.245 | -0.071 | 0.292 | -0.071 | 0.419 | -0.058 | 0.46 | -0.056 |
| Suivant Y | -0.107 | -0.186 | -0.113 | -0.186 | -0.097 | -0.206 | -0.1 | -0.216 |
| Suivant Z | -0.964 | 0.98 | -0.95 | 0.98 | -0.903 | 0.977 | -0.882 | 0.975 |

De plus il existe une bonne corrélation entre ces deux plans (déplacements, efforts) puisque l'angle entre les normales $\vec{n}_f$ et $\vec{n}_u$ ne dépasse pas 2,5° tableau 2.

**Tableau 2.** Corrélation entre le plan des efforts et le plan des déplacements.

| f (mm/tr) | $arc\cos\left(\dfrac{\vec{n}_f \cdot \vec{n}_u}{\|\vec{n}_f\| \cdot \|\vec{n}_u\|}\right)$ |
|---|---|
| 0.05 | 0.8° |
| 0.0625 | 1° |
| 0.075 | 1.9° |
| 0.1 | 2.5° |

En analysant le plan des déplacements ($P_u$), nous observons dans la figure 11 la concordance entre l'angle d'inclinaison de la trace de ce plan ($\delta_u$) dans le plan de coordonnées YZ (dont l'évolution reste faible en fonction de l'avance) et l'angle ($\delta_{ozy} = 16°$) de la raideur équivalente projetée dans ce même plan de coordonnées YZ (Fig. 4).

Ces résultats sont en accord avec [6] à [9] et montrent que la composante dans la direction du mouvement vibratoire de la structure élastique de la machine-outil provoque des variations de section du copeau, conséquence des vibrations auto-entretenues.

Ces relations mettent en évidence le couplage entre les caractéristiques élastiques du système **BO** et les vibrations générées par la coupe. Comme nous pouvions nous y attendre, l'apparition des vibrations auto-entretenues est fortement influencée par les valeurs des raideurs du système, leur rapport et leur direction.

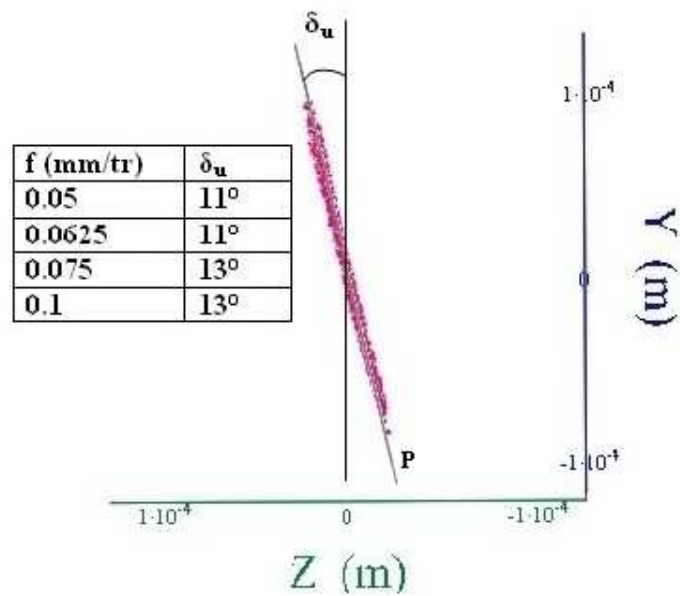

**Fig. 11.** Faible évolution de l'inclinaison de la trace du plan des déplacements dans le plan de coordonnées YZ en fonction de l'avance.

L'existence et la détermination de ce plan des déplacements sont essentielles dans la modélisation du comportement dynamique de la coupe. La détermination de ce plan fournit des informations sur la configuration à adopter pour l'écriture du modèle et permet d'exprimer le système en fonction de ces axes. Ainsi, le caractère plan du comportement du **BO** permet de ramener le problème tridimensionnel de coupe, avec déplacement spatial, à une modélisation dans un plan incliné par rapport aux axes de la machine. Néanmoins, cela reste un problème de coupe tridimensionnel [2].

Grâce à l'identification de ces plans (plans des déplacements, plan des efforts), nous pouvons déterminer par la suite les configurations réelles de coupe et proposer un modèle dynamique intégrant la géométrie de l'outil et le contact pièce / outil / copeau.

**6 Conséquences sur la modélisation**

Les efforts estimés à l'aide du modèle de coupe sont à la base du modèle dynamique en cours de développement au sein des laboratoires bordelais [10] et prenant en compte les variations de contact outil / copeau et outil / pièce.

Le comportement dynamique de la coupe est décrit de façon générale par :

$$[M] \cdot \ddot{\vec{u}} + [C] \cdot \dot{\vec{u}} + [K] \cdot \vec{u} = L_{CT}(Y_{cs}(t) \cdot \vec{T_{OB}} + \vec{T_{OJ}} + L(t) \cdot \vec{T_{JK}}) \qquad (5)$$

où [M] représente la matrice de masse, [C] la matrice d'amortissement, [K] la matrice de raideur, et $\vec{u}$ le vecteur position. Dans ce système, la force dynamique est fonction des paramètres du contact de l'outil : $Y_{cs}$ la longueur de la zone de cisaillement secondaire OB, $\vec{T_{OB}}$ le vecteur contrainte sur cette même zone OB, $\vec{T_{OJ}}$ le vecteur contrainte sur le rayon d'arête de l'outil, L la longueur de contact de la zone de dépouille, $\vec{T_{JK}}$ le vecteur contrainte dans la zone JK, $L_{CT}$ la longueur d'arête mesurée suivant la profondeur de passe (c'est-à-dire l'axe X).

Sachant que les déplacements de l'outil ont lieu dans un plan connu grâce aux résultats des paragraphes précédents (Fig. 10), il est alors possible de simplifier la modélisation tridimensionnelle précédente en un système de deux équations différentielles à deux degrés de liberté dans ce plan (Fig. 12).

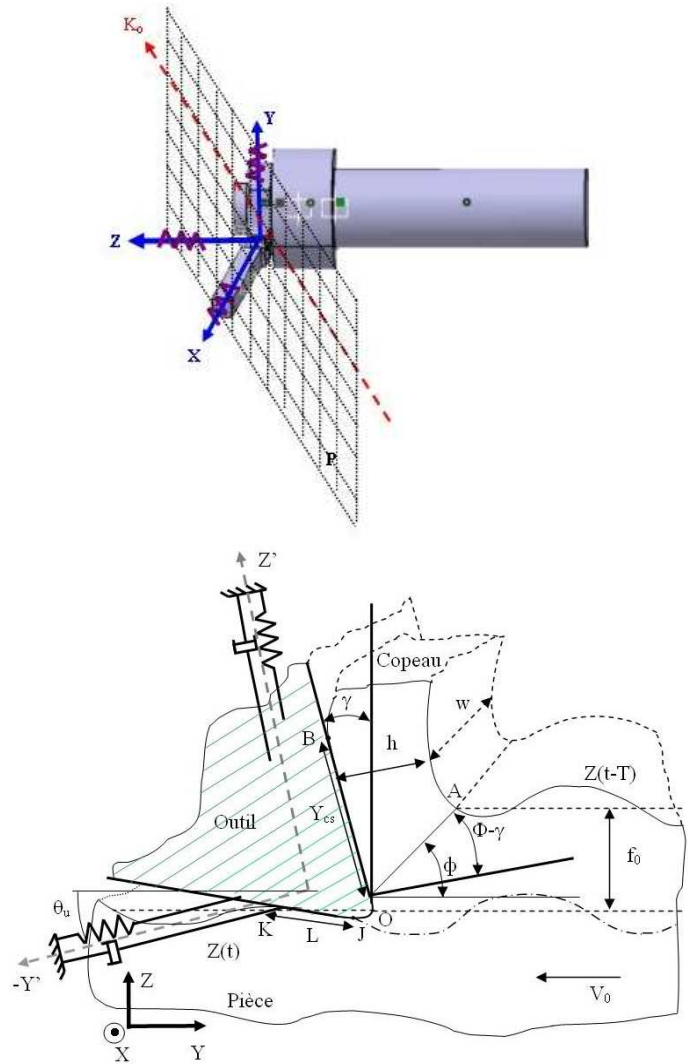

**Fig. 12.** Description du modèle dynamique dans le repère lié au plan des déplacements.

## 7 Conclusions et perspectives

Les procédures expérimentales mises en place, tant au niveau statique que dynamique, ont permis de déterminer les éléments nécessaires à une analyse rigoureuse de l'influence de la géométrie de l'outil, de son déplacement et de l'évolution des contacts outil / pièce et outil / copeau sur la surface réalisée.

Le couplage mis en évidence entre les caractéristiques élastiques du système **BO** et les vibrations générées par la coupe ont permis d'établir que l'apparition des vibrations auto-entretenues est fortement influencée par les raideurs du système, leur rapport et leur direction.

De plus, les phénomènes de vibrations auto-entretenues ont clairement été isolés indépendamment du système usinant (**BO** et **BP**).

L'ensemble des résultats expérimentaux obtenus a conduit à la simplification du modèle tridimensionnel vers un modèle à 2 degrés de liberté dans le plan des déplacements identifié lors de cette étude.

Ces premiers résultats permettent d'envisager à présent une étude plus complète en exploitant totalement la notion de torseur. Ainsi, les différents couplages [$K_{icf}$], [$K_{ifc}$] sur les raideurs [$K_i$] non pris en compte dans cette étude, entre le torseur d'actions mécaniques (forces et moments) et le torseur des déplacements (translations et rotations) seront étudiés ultérieurement. Ainsi, grâce au dynamomètre à six composantes, nous avons identifié l'existence de moments à la pointe de l'outil [11]. L'exploitation de l'évolution de ceux-ci (liés aux matrices [$K_{ic}$], [$K_{icf}$], [$K_{ifc}$]) est en cours et devrait prochainement compléter harmonieusement ce travail.

**Références**